\documentclass[aps,twocolumn,showpacs,floatfix,letterpaper]{revtex4}
\pdfoutput=1
\usepackage[latin1]{inputenc}
\usepackage[T1]{fontenc}
\usepackage[english]{babel}
\usepackage{times}
\usepackage{graphicx}
\usepackage{amssymb}

\usepackage{color}

\usepackage[sort&compress]{natbib} 
\usepackage{hypernat} 
\usepackage[final,colorlinks=true,urlcolor=blue,citecolor=blue,linkcolor=blue]{hyperref}
\hypersetup{pdfauthor   = {Guillaume Lang}}

\newcommand{\etal}[0]{\textit{et al.}}
\newcommand{\LDL}[0]{LD-like}           
\newcommand{\HDL}[0]{HD-like}           
\newcommand{\xNQR}[0]{\ensuremath{x_{NQR}}}
\newcommand{\xFSWT}[0]{\ensuremath{x_{FSWT}}}
\newcommand{\xNQRxFSWT}[0]{\ensuremath{\xNQR / \xFSWT}}
\newcommand{\VSC}[0]{\ensuremath{V_{SC}}}
\newcommand{\DnuDT}[0]{\ensuremath{\Delta \nu_Q / \Delta T}}

\begin{document}
\title{Spatial competition of the ground states in 1111 iron pnictides}
\author{G. Lang,\textsuperscript{1,2,*} L. Veyrat,\textsuperscript{1} U. Gr\"afe,\textsuperscript{1} F. Hammerath,\textsuperscript{1,3}  D. Paar,\textsuperscript{1,4} G. Behr,\textsuperscript{1} S. Wurmehl,\textsuperscript{1,3} and H.-J. Grafe\textsuperscript{1}}
\affiliation{
\textsuperscript{1} IFW Dresden, Institute for Solid State Research, PF 270116, D-01171 Dresden, Germany \\
\textsuperscript{2} ESPCI ParisTech, PSL Research University; CNRS; Sorbonne Universit\'es, UPMC Univ. Paris 6; LPEM, 10 rue Vauquelin, F-75231 Paris Cedex 5, France \\
\textsuperscript{3} Institut f\"{u}r Festk\"{o}rperphysik, Technische Universit\"{a}t Dresden, D-01062 Dresden, Germany \\
\textsuperscript{4} Department of Physics, Faculty of Science, University of Zagreb, P.O. Box 331, HR-10002 Zagreb, Croatia}

\date{\today}

\begin{abstract}
Using nuclear quadrupole resonance, the phase diagram of 1111 $R$FeAsO$_{1-x}$F$_x$ ($R$$=$La, Ce, Sm) iron pnictides is constructed as a function of the local charge distribution in the paramagnetic state, which features low-doping-like (LD-like) and high-doping-like (HD-like) regions.
Compounds based on magnetic rare earths (Ce, Sm) display a unified behavior, and comparison with La-based compounds reveals the detrimental role of static iron $3d$ magnetism on superconductivity, as well as a qualitatively different evolution of the latter at high doping. 
It is found that the LD-like regions fully account for the orthorhombicity of the system, and are thus the origin of any static iron magnetism.
Orthorhombicity and static magnetism are not hindered by superconductivity but limited by dilution effects, in agreement with 2D (respectively 3D) nearest-neighbor square lattice site percolation when the rare earth is nonmagnetic (respectively magnetic).
The LD-like regions are not intrinsically supportive of superconductivity, on the contrary of the HD-like regions, as evidenced by the well-defined Uemura relation between the superconducting transition temperature and the superfluid density when accounting for the proximity effect. 
This leads us to propose a complete description of the interplay of ground states in 1111 pnictides, where nanoscopic regions compete to establish the ground state through suppression of superconductivity by static magnetism, and extension of superconductivity by proximity effect.
\end{abstract}


\maketitle

\section{Introduction}

One of the main motivations to study iron-based high-temperature superconductors (IBS) is the frequent vicinity of unconventional superconductivity and static magnetism in their phase diagram.
Also observed in copper-based superconductors, it raises the questions of ground state interplay and of the role of spin fluctuations in the Cooper pairing interaction.
By applying pressure, changing the doping or altering the electronic structure via isovalent substitutions \cite{Stewart2011,Wang2009k}, the ground state of most IBS can be modified from static magnetism towards superconductivity, with the possibility of ground state coexistence on a microscopic scale in several cases \cite{Drew2009,Sanna2009,Carlo2009,Shiroka2011,Lamura2015,Laplace2009,Julien2009,Wiesenmayer2011}.

A key aspect of iron-based superconductors is the itinerant, multi-band character of their electronic structure.
In compounds such as the 1111 and 122 families, the low-temperature magnetic order of the parent compounds ($T_N\approx 140~K$ for the 1111 family) is argued to be a spin-density wave caused by the nesting properties of the Fermi surface \cite{Korshunov2008a,Cruz2008}.
As this transition is associated to a tetragonal-to-orthorhombic structural transition at the same or a slightly-higher temperature, both Ising-nematic spin fluctuations and orbital fluctuations have been invoked to explain the observed anisotropic electronic response of the iron plane above the magnetic transition \cite{Fernandes2014}.
Evolutions of the Fermi surface with doping would affect the competition of ground states, and relative changes in interband and intraband scattering would play a role in determining whether superconductivity is associated to spin or orbital fluctuations, yielding respectively a $s\pm$ or $s_{++}$ symmetry of the order parameter \cite{Chubukov2008,Mazin2008,Kuroki2008,Kontani2010}.

The ground state competition is, however, complicated by spatial electronic inhomogeneities which derive either from the local effect of in-plane chemical substitutions \cite{Texier2012a,Laplace2012,Hammerath2014}, or from intrinsic iron plane physics.
Using Nuclear Quadrupole Resonance (NQR), a local probe of the charge environment, we have previously shown that the fluorine-doped 1111 family ($R$FeAsO$_{1-x}$F$_x$ with $R$ a rare earth) features two types of local electronic environments intrinsic to the FeAs layers when moderately doped \cite{Lang2010}.
Although these environments are already defined in the paramagnetic state, their simultaneous presence in purely magnetic or purely superconducting samples appears to rule out a trivial explanation of any ground state coexistence in 1111 compounds.
Further complicating the analysis of ground state interplay, such coexistence seems dependent on the choice of rare earth (La, Ce, Sm\ldots) \cite{Luetkens2009,Drew2009}.
Obscuring the whole phase diagram is the difficulty in assessing the true doping, which may differ dramatically from the nominal doping depending on the rare earth and the synthesis route  \cite{Koehler2009,Rotundu2009,Fujioka2013a}.

In this article, we present how the nanoscale charge environments observed by NQR in the paramagnetic state yield a highly-accurate determination of the effective doping, allowing us to compare unambiguously the phase diagrams of $R$FeAsO$_{1-x}$F$_x$ ($R$$=$La, Ce, Sm) and to explain the evolution of ground state properties with doping.
Sec.~\ref{str:phasediag} describes how to obtain such phase diagrams, which are found undistinguishable for the two magnetic rare earths (Ce, Sm).
The different behavior for lanthanum shows that static magnetism from the iron has a detrimental effect on superconductivity, while at high doping La-based samples display a specific, nonmonotonous behavior of the superconducting transition temperature.
In Sec.~\ref{str:magnetism}, it is found that the low-doping-like regions seen by NQR fully account for the orthorhombicity and iron static magnetism of the system.
Nearest-neighbor square lattice percolation is at play, with dimensionality being dependent on whether the rare earth itself is magnetic.
In Sec.~\ref{str:SC}, it is shown that the low-doping-like regions can only host superconductivity by proximity, with intrinsic superconductivity originating from the high-doping-like regions seen by NQR.
Finally, we propose in Sec.~\ref{str:interplay} a full description of the interplay of ground states in 1111 pnictides, where superconductivity extends spatially whenever static magnetism is weak enough in the nearby microscopic regions.

\section{Experimental details}
\label{str:exp_details}

Two different routes were followed for the synthesis of the polycrystalline $R$FeAsO$_{1-x}$F$_x$ ($R$$=$La, Ce, Sm) samples.
For route 1, FeAs was prepared by a solid-state reaction prior to the synthesis of the corresponding 1111 compounds. Subsequently, the resulting FeAs was mixed with metallic $R$, $R_2$O$_3$, and $R$F$_3$ in a stoichiometric ratio. This mixture was homogenized by grinding in a mortar. 
For route 2, we prepared $R$As as first step by reacting $R$ and As lumps in a stoichiometric ratio via a vapor transport reaction. The second step of route 2 used the resulting $R$As, Fe, Fe$_2$O$_3$, and FeF$_3$ as starting materials in a stoichiometric ratio. Here, the starting materials were homogenized by grinding in a ball mill.
In either case, the resulting powders were pressed into pellets under Ar atmosphere, and subsequently annealed in an evacuated quartz tube in a two-step synthesis first at 940$^\circ$C for 12~h and then at 1150$^\circ$C for up to 60~h.
All samples were characterized by powder X-ray diffraction (XRD) and by scanning electron microscopy (SEM) with semiquantitative elemental analysis using the wavelength dispersive X-ray (WDX) mode.
Their structural properties and their structural, magnetic, and superconducting phase transitions were determined using a wide array of techniques: standard XRD, synchrotron XRD at low temperature, magnetic susceptibility, electrical resistivity, muon spin relaxation and rotation ($\mu$SR), M\"{o}ssbauer spectroscopy, and nuclear magnetic resonance (NMR) \cite{Luetkens2009,Kondrat2009,Hess2009,Klingeler2010,Maeter2012,Hammerath2013}.
The determination of the doping will be discussed in the next Section, with all phase transitions being included in Fig.~\ref{fig:phase_diag_NQR}.

In order to probe the distribution of charge environments at the nanometer scale, $^{75}$As nuclear quadrupole resonance (NQR) was used.
NQR probes nuclei with spin $I$$>$1/2, which possess a finite electric quadrupole moment $eQ$ with $e$ the elementary charge and $Q$ the nuclear quadrupole moment.
In the absence of applied or internal magnetic field, any finite electric field gradient (EFG) at the atomic site will lift at least partially the degeneracy of the nuclear electric quadrupole energy levels, according to the following Hamiltonian \cite{Abragam1961}:
\begin{equation}
 {\cal H} = \frac{eQ V_{zz}}{4I(2I-1)} \left[3 I_{z}^2 - I(I+1) + \frac{\eta}{2}(I_+^2+I_-^2)\right]  \mbox{,}
 \label{eq:hamiltonian}
\end{equation}
whose coordinate axes are defined by the principal axes of the electric field gradient (EFG) tensor, itself characterized by its largest principal value $V_{zz}$ and its asymmetry parameter $\eta$$=$$(V_{xx}-V_{yy})/V_{zz}$ ($|V_{zz}|\ge|V_{yy}|\ge|V_{xx}|$ and 0$\le \eta \le$1).
Since $^{75}$As has a nuclear spin $I$$=$3/2, the Hamiltonian yields a single resonance frequency:
\begin{equation}
 \nu_Q = \frac{eQ V_{zz}}{2h} \sqrt{1+\frac{\eta^2}{3}}   \mbox{,}
 \label{eq:nuQ}
\end{equation}
where $h$ is Planck's constant.
The electric quadrupole frequency $\nu_Q$ depends on the symmetry and amplitude of the EFG, which itself depends on the atomic/nanoscale charge environment.
For every local charge environment as seen from the arsenic sites, one resonance line will be observed.

Using a standard pulsed NQR spectrometer, radiofrequency irradiations over a typical $\nu_{rf}$$=$8--14~MHz frequency range were performed to obtain the histogram of resonance frequencies for all arsenic sites in the sample.
Powder samples were crushed to ensure that the crystallites are small enough to obtain a good penetration of the radiofrequency field.
Note that operating on powders does not affect the spectra, since $\cal H$ depends on the principal axes of the local EFG, i.e., the resonance frequency for a given charge environment is independent of the orientation of individual crystallites.
Spin echo sequences ($\frac{\pi}{2}$--$\tau$--$\pi$) were used, with typical $\tau$$=$20--30~$\mu$s.
The $\tau$ values and the pulse sequence repetition rates were checked to be small enough that spin-spin or spin-lattice relaxation contrast do not distort the relative line intensities.
Point-by-point spectra were obtained by integrating the full echo and applying a $\nu_{rf}^{-2}$ intensity correction.

\section{Reconstructed phase diagram}
\label{str:phasediag}

In most of the published literature, nominal dopings are used to build the phase diagrams of iron-based superconductors.
While this yields qualitatively correct results for a given sample series, it is known that the real doping may be significantly lower, especially for high nominal dopings \cite{Koehler2009,Rotundu2009,Fujioka2013a}.
It is thus difficult to extract quantitative information (e.g.: doping thresholds), and to compare different compounds in terms of ground-state competition and the doping profile of the superconductivity dome.
For F-doped 1111 compounds, real content determination using WDX spectroscopy is hardly possible for cerium-based compounds due to the superposition of the relevant Ce and F lines, and may generally be affected by the presence of poorly-crystallized impurities which XRD cannot easily account for.
Using fluorine NMR, the doping could be established within an absolute error of 2\% \cite{Sanna2010,Shiroka2011}, which is satisfactory for high dopings but may be imprecise to discuss the boundary between the two ground states at low doping.
Here, we propose to take advantage of the doping dependence of the charge environments previously seen by NQR for La- and Sm-based samples \cite{Lang2010}.

\begin{figure*}[htbp]
\includegraphics[width=120mm]{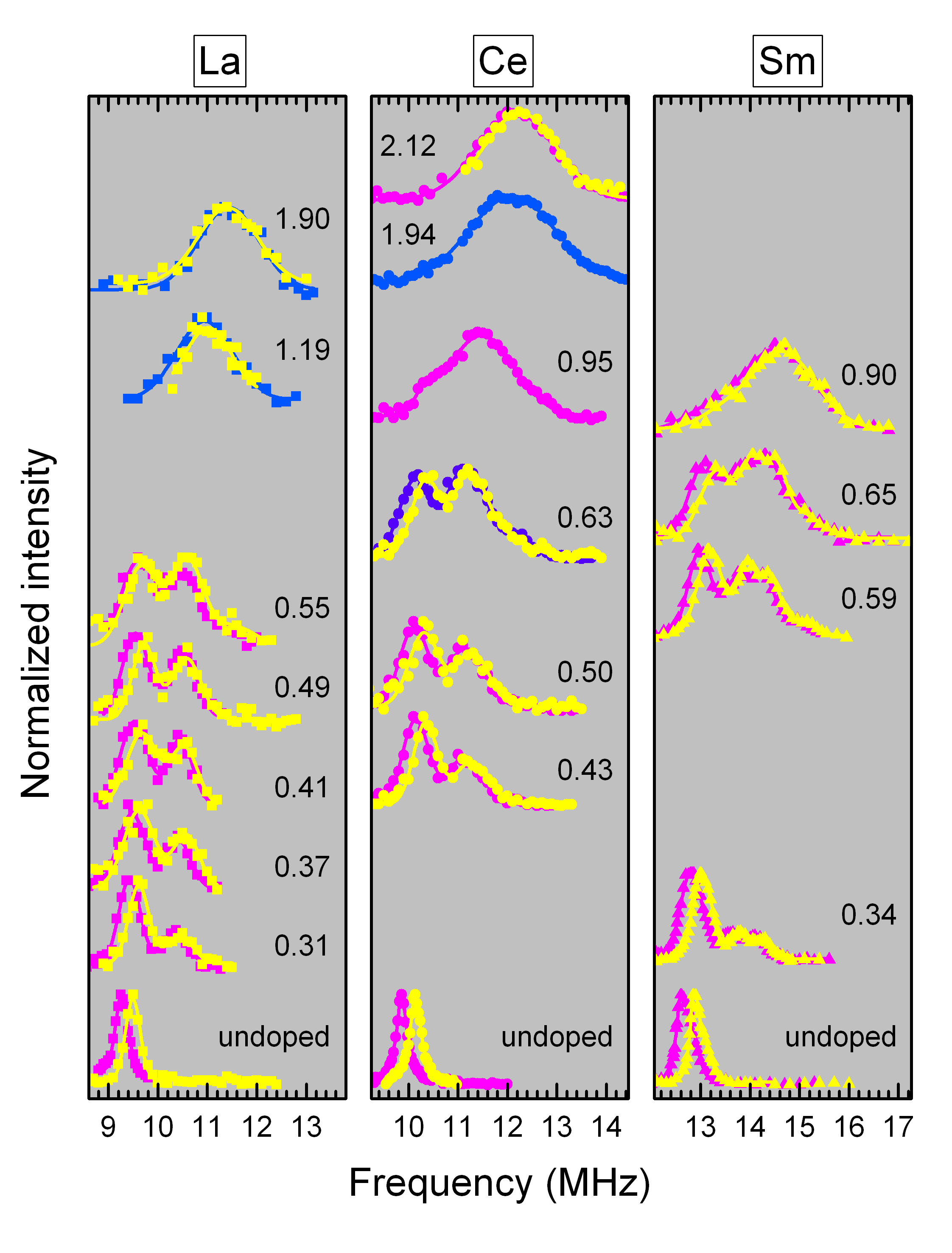}
\caption{(Color online) NQR spectra of the studied samples in the paramagnetic state. The doping is indicated as \xNQR$/$\xFSWT\ (FSWT: full spectral weight transfer, see text). The measurement temperature is coded by the color (see also the scale in Fig.~\ref{fig:nuQvsXnqr}): Yellow for room temperature, magenta for 150~K to 160~K, indigo for 100~K (Ce, \xNQR$/$\xFSWT$=$0.63), blue for 40~K to 55~K.
Full lines are Gaussian fits. Some low-temperature La/Sm spectra are from a previous study \cite{Lang2010}.}
\label{fig:spectra}
\end{figure*}

\subsection{Principle}
\label{str:phase_diag_principle}

Fig.~\ref{fig:spectra} shows the NQR spectra for $R$-1111 ($R$ $=$ La, Ce, Sm) in the paramagnetic state over a broad doping range (itself determined by NQR as described below, see also Tab.~\ref{tab:NQR} for the fluorine content). 
Whereas undoped and highly-doped samples display a single peak corresponding to a single charge environment, the samples at intermediate doping display two peaks (two charge environments) which were shown to correspond to a nanoscale electronic inhomogeneity intrinsic to the FeAs layers \cite{Lang2010}.
Defining $\nu_Q^L$ ($\nu_Q^H$) and $w^L$ ($w^H$) as the low-frequency (high-frequency) peak's position and spectral weight ($w^L$$+$$w^H$$=$1), and $x$ as the WDX-established doping, these two charge environments were found to have the following properties:
(i) the transfer of spectral weight is linear with $x$ and is completed ($w^L = 0$) at $x$$\approx$0.1
(ii) $\nu_Q^L$ and $\nu_Q^H$ vary linearly with $x$, extrapolating respectively to the frequencies of the $x$$=$0 and $x$$=$0.1 charge environments
(iii) the linear $x$ dependence of $\nu_Q^H$ appeared to extend beyond $x$$=$0.1, which was independently confirmed \cite{Oka2011}.
The low- and high-frequency peaks were thus associated to low-doping-like (\LDL) and high-doping-like (\HDL) regions.

\begin{figure*}[pbth]
\includegraphics[width=165mm]{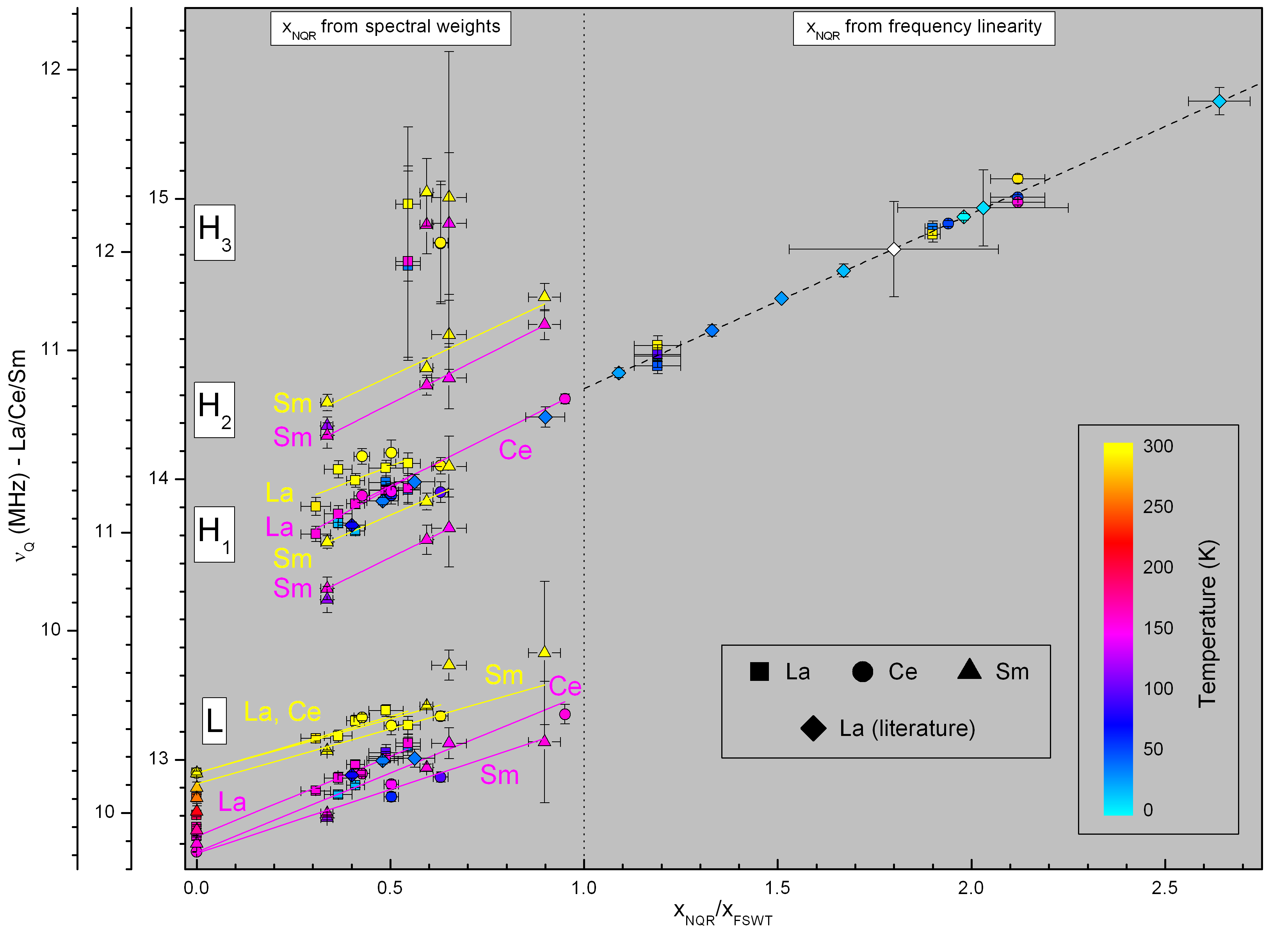}
\caption{(Color online) Quadrupolar frequency versus \xNQR$/$\xFSWT. The L/H$_1$/H$_2$/H$_3$ labels refer to the observed frequency branches. Literature data is taken from Refs.~\onlinecite{Kitagawa2010c,Oka2011,Tatsumi2009,Mukuda2010,Kobayashi2010,Nitta2012} (for one sample, the unknown measurement temperature is color-coded as white). For \xNQR$/$\xFSWT$<$1, \xNQR\ is determined using the NQR spectral weights (see text for comments about some samples from Ref.~\onlinecite{Oka2011}). Linear fits are indicated as full lines with attached labels. For each sample, the error on \xNQR\ is derived from fits of the spectra at one or more temperatures. For \xNQR$/$\xFSWT$>$1, \xNQR\ is determined from the linear extrapolation of the low \xNQR\ behavior, indicated as a dashed line. For each sample, the error on \xNQR\ is set so as to account for the uncertainty on $\nu_Q$, whether experimental on a single point or due to remaining temperature dependence.}
\label{fig:nuQvsXnqr}
\end{figure*}

\begin{table}
\begin{tabular}[c]{ccccc}
\hline
\hline
$R$	&	Source					& 	$x$	&	\xNQRxFSWT	& \xFSWT	\\
\hline

La      &       \textit{this study}                             &       0       &       0.00\phantom{()11}      &               \\ 
        &                                                       &       0.035   &       0.31(4)\phantom{1}      &       0.11    \\ 
        &                                                       &       0.04    &       0.37(4)\phantom{1}      &       0.11    \\ 
        &                                                       &       0.045   &       0.41(2)\phantom{1}      &       0.11    \\ 
        &                                                       &       0.05    &       0.49(4)\phantom{1}      &       0.10    \\ 
        &                                                       &       0.075   &       0.55(3)\phantom{1}      &       0.14    \\ 
        &                                                       &       0.1     &       1.19(6)\phantom{1}      &       0.08    \\ 
        &                                                       &       0.15    &       1.90(2)\phantom{1}      &       0.08    \\ 
 & & & & \\ 
Ce      &       \textit{this study}                             &       0       &       0.00\phantom{()11}      &               \\ 
        &                                                       &       0.05    &       0.43(2)\phantom{1}      &       0.12    \\ 
        &                                                       &       0.15    &       0.50(2)\phantom{1}      &       0.30    \\ 
        &                                                       &       0.1     &       0.63(2)\phantom{1}      &       0.16    \\ 
        &                                                       &       0.1     &       0.95\phantom{()11}      &       0.11    \\ 
        &                                                       &       0.2     &       1.94\phantom{()11}      &       0.10    \\ 
        &                                                       &       0.25    &       2.12(7)\phantom{1}      &       0.12    \\ 
 & & & & \\ 
Sm      &       \textit{this study}                             &       0       &       0.00\phantom{()11}      &               \\ 
        &                                                       &       0.04    &       0.34(2)\phantom{1}      &       0.12    \\ 
        &                                                       &       0.06    &       0.59(2)\phantom{1}      &       0.10    \\ 
        &                                                       &       0.08    &       0.65(4)\phantom{1}      &       0.12    \\ 
        &                                                       &       0.1     &       0.90(4)\phantom{1}      &       0.11    \\ 
 & & & & \\ 
La      &       \cite{Kitagawa2010c} - Kitagawa &       0.14    &       1.98\phantom{()11}      &       0.07    \\ 
        &       \cite{Oka2011} - Oka    &       0.03    &       0.40(4)\phantom{1}      &       0.07    \\ 
        &                                                       &       0.04    &       0.48(4)\phantom{1}      &       0.08    \\ 
        &                                                       &       0.06    &       0.90(5)\phantom{1}      &       0.07    \\ 
        &                                                       &       0.08    &       1.09\phantom{()11}      &       0.07    \\ 
        &                                                       &       0.1     &       1.67\phantom{()11}      &       0.06    \\ 
        &                                                       &       0.15    &       2.64(8)\phantom{1}      &       0.06    \\ 
        &       \cite{Tatsumi2009} - Tatsumi    &       0.14    &       1.51\phantom{()11}      &       0.09    \\ 
        &       \cite{Mukuda2010} - Mukuda      &       0.22    &       2.03(22)        &       0.11    \\ 
        &       \cite{Kobayashi2010} - Kobayashi        &       0.11    &       0.56(5)\phantom{1}      &       0.20    \\ 
        &                                                       &       0.15    &       1.33\phantom{()11}      &       0.11    \\ 
        &       \cite{Nitta2012} - Nitta        &       0.14    &       1.80(27)        &       0.08    \\

\hline
\hline
\end{tabular}
\caption{Doping information on the samples from this study, as well as on samples from previous LaO$_{1-x}$F$_x$FeAs NQR/NMR studies whose spectral properties have been re-analyzed following the same procedure as for our samples. $R$ indicates the rare earth. For samples from the literature, the name of the first author is given next to the reference number. $x$ indicates the fluorine doping. For our La-based samples it is the nominal value, which is on the order of the WDX value, whereas for our Sm-based samples it is a rescaled value derived from WDX measurements \cite{Kondrat2009,Hess2009}. For our Ce samples it is the nominal value, without the possibility of comparison to WDX values due to the superposition of the relevant Ce and F lines.
For samples from the literature it is usually the nominal value.
\xNQRxFSWT\ is the NQR-derived doping as defined in Sec.~\ref{str:phase_diag_principle}.
\xFSWT\ is the per-sample extrapolated fluorine content for which full spectral weight transfer would occur in the NQR spectrum (see Sec.~\ref{str:phase_diag_xFSWT}).
}
\label{tab:NQR}
\end{table}

Taking advantage of the fact that NQR is immune to spurious phases and allows to measure with good precision the peak positions and the spectral weights in practically all samples,  we reverse the above observations to define \xNQR, a NQR-defined doping.
For samples with \textit{doping low-enough that two charge environments are observed}, following point (i) we set:
\begin{equation}
  \xNQR = w^H \xFSWT   \mbox{,}
\end{equation}
with \xFSWT\ the value for which full spectral weight transfer (FSWT) has occurred, i.e., the electronic homogeneity is restored.
For samples with \textit{doping large enough that a single charge environment is observed} ($w^H$$=$1), following point (iii) we set \xNQR\ such that $\nu_Q^H$ is linear versus this variable.

While it seems from point (i) that \xFSWT\ corresponds to about 10\% fluorine content, we will show in the following that this value cannot be estimated very precisely and that it is more reliable to express \xNQR\ in units of \xFSWT\ (i.e., to determine the ratio \xNQRxFSWT) rather than as a fluorine-content equivalent doping.

\subsection{Data extraction}

From all measured spectra and using our rules for determining \xNQRxFSWT, we obtain Fig.~\ref{fig:nuQvsXnqr}, where the vertical axes are offset so that there is overlap for all three rare earths at \xNQR$=$0 and room temperature.
Such overlap is preserved for the whole $\nu_Q^L$ low-frequency branch (``L'') at room temperature, with a well-defined linear dependence on \xNQR\ (yellow fits) as expected from the above points (i) and (ii).
This linear behavior holds against changes in temperature, albeit with different slopes for the different rare earths (purple fits for $T$$\approx$160~K).
The influence of temperature will be considered when discussing orthorhombicity and static magnetism in Sec.~\ref{str:magnetism}.
For the $\nu_Q^H$ high-frequency branch with \xNQR$<$\xFSWT, the situation is complicated by the presence of a superstructure in the high-frequency peak of the Sm samples, which appears to be partially echoed in the \xNQRxFSWT$=$0.55 La sample and \xNQRxFSWT$=$0.63 Ce sample, thus a total of three high-frequency branches (``H$_1$/H$_2/$H$_3$'').
Nonetheless, similar linear behaviors account well for the data (see yellow and purple fits).
For the samples with \xNQR$>$\xFSWT, i.e., only La- or Ce- based samples, we apply the above procedure of extrapolating the linear behavior at \xNQR$<$\xFSWT.
The H$_1$ frequency branch is used, since the H$_2$ and H$_3$ branches appear to play little to no role for La- and Ce-based samples.
Note that this may not hold true for Sm-based samples (no sufficiently highly-doped samples were available to test for this).
Since at high doping the temperature dependence of the spectra vanishes (La) or is much reduced (Ce), we use a single extrapolation intermediate between the low-temperature (La, Ce) and room-temperature (La) H$_1$ fits (no reliable room-temperature H$_1$ fit is available for Ce, but the corresponding data is roughly similar to the La data).
This interpolation is shown as a dashed line on Fig.~\ref{fig:nuQvsXnqr}.

The procedure to determine \xNQRxFSWT\ was also applied to LaO$_{1-x}$F$_x$FeAs NQR/NMR data from the literature (Refs.~\onlinecite{Kitagawa2010c,Oka2011,Tatsumi2009,Mukuda2010,Kobayashi2010,Nitta2012}).
All included samples are listed in Tab.~\ref{tab:NQR} together with their \xNQRxFSWT\ value.
In the following, they are designated as ``source-$x$'', where source is the first author of the original paper and $x$ the reported fluorine content.
In most cases, the availability of the NQR spectra allowed us to re-analyse them in the same way as for our own samples.
A few samples for which data was too sparse or fit anomalies were present are excluded.
For the samples with two-peaked spectra (Oka-0.03, Oka-0.04, and Kobayashi-0.11), the spectral weights obtained from the fits yield \xNQRxFSWT\ values respectively equal to 0.40(4), 0.57(2), and 0.56(5).
However, the reliability of these values may be affected by the experimental conditions under which the spectra were measured, i.e., the radiofrequency pulse sequence repetition time and the delay $\tau$ between the pulses (see Sec.~\ref{str:exp_details}).
If the former is too short or the latter too long, the relative spectral weights will be modified due to differences in $T_1$ (spin-lattice relaxation time) or $T_2$ (spin-spin relaxation time) between the two peaks \cite{Lang2010}.
This may be the case for the Oka-0.06 sample, whose single NQR peak's frequency implies \xNQRxFSWT$=$0.90(5), i.e., a value for which two peaks should be observed.
Here, the expected low-frequency peak may be entirely missed due to its small weight (10(5)\%) being further reduced by so-called $T_2$ relaxation contrast.
Considering the resonance frequencies obtained from the fits, the above \xNQRxFSWT\ values for Oka-0.03 and Kobayashi-0.11 appear reasonably compatible with the L and H$_1$ frequency branches defined by our samples (see the corresponding diamond symbols on Fig.~\ref{fig:nuQvsXnqr}).
For Oka-0.04, the agreement is somewhat poorer and suggests that the actual \xNQRxFSWT\ value is lower than the fit-derived value of 0.57(2). 
This is supported by the comparison of spin-lattice relaxation rate measurements in Ref.~\onlinecite{Oka2011} and Ref.~\onlinecite{Lang2010}, showing that Oka-0.04 is indeed rather lower doped than our 5\% fluorine sample for which \xNQRxFSWT$=$0.49(4).
Based on agreement with the L and H$_1$ frequency branches, \xNQRxFSWT\ is then estimated to be 0.48(4), as used in Fig.~\ref{fig:nuQvsXnqr}.

\subsection{Estimation of \xFSWT}
\label{str:phase_diag_xFSWT}

All obtained \xNQRxFSWT\ values are reported in Tab.~\ref{tab:NQR}.
To get an estimate of \xFSWT, one can compute $x$$(\xNQRxFSWT)^{-1}$ for every sample: by setting \xNQR$=$$x$, \xFSWT\ will be the fluorine content for which full NQR spectral weight transfer would occur in a given sample, \textit{with an inaccuracy corresponding to that present in $x$}.
As can be seen in Tab.~\ref{tab:NQR} most computed values of \xFSWT\ are in the range 0.06--0.12, while a few values are slightly (0.14, 0.16) or much (0.20, 0.30) higher.
The most likely candidate for this variability is fluorine that did not enter the matrix 1111 phase (pushing \xFSWT\ up for a given sample), although other sources cannot be ruled out (e.g.: small amounts of oxygen or arsenic vacancies which might drive down \xFSWT).
This variability plays however no role in our study, where all deductions will be made on the basis of the \xNQRxFSWT\ ratio, which does not depend on any chemical characterization.

\begin{figure*}[htbp]
\includegraphics[width=150mm]{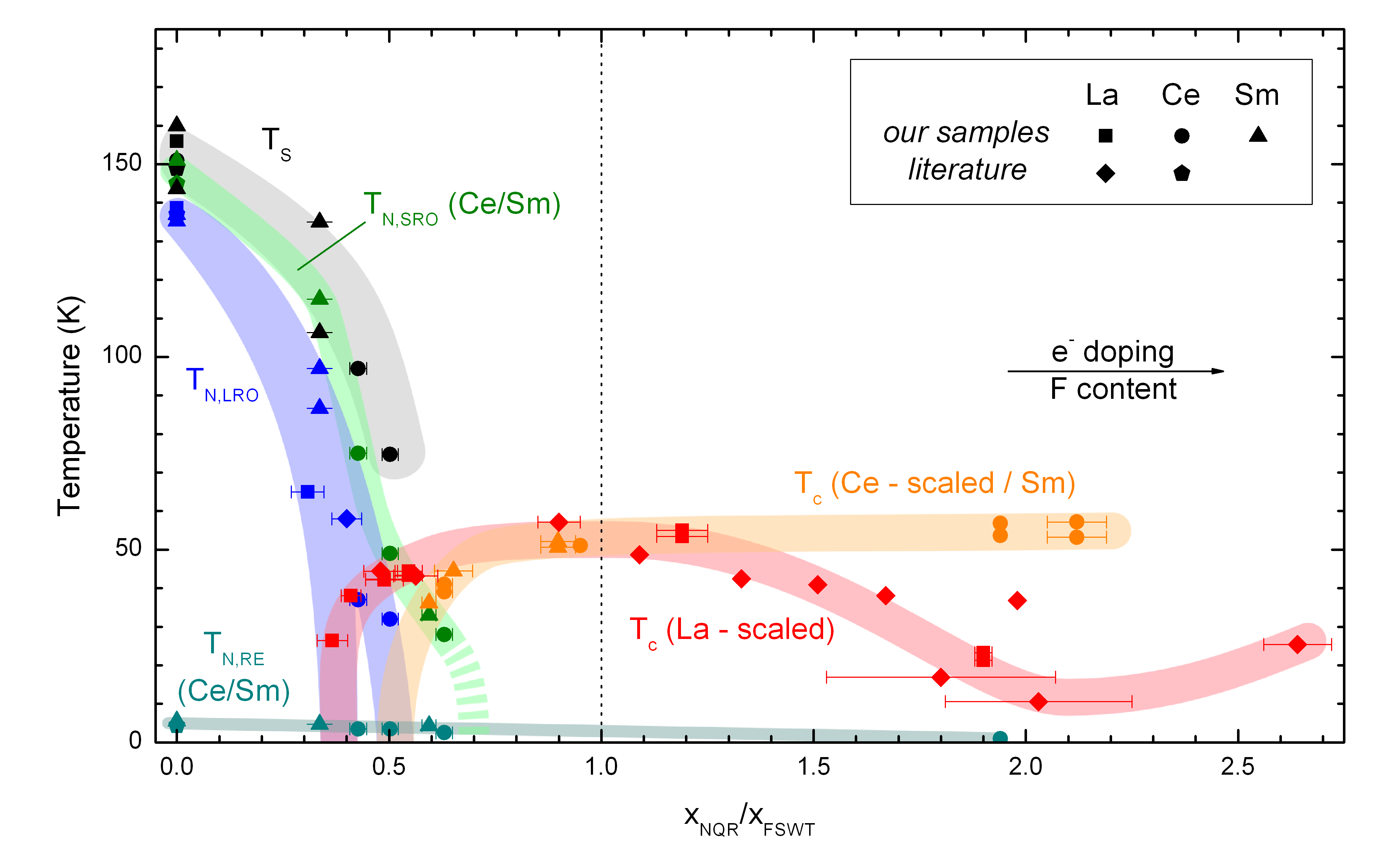}
\caption{(Color online) Phase diagram of $R$FeAsO$_{1-x}$F$_x$ ($R$$=$La, Ce, Sm) as a function of \xNQRxFSWT. $T_c$ data for La- and Ce-based samples have been scaled (see text). For some samples and some transitions, more than one point may be displayed due to characterization using multiple techniques. The samples from the literature are referenced in Tab.~\ref{tab:NQR}, except the undoped Ce-based sample from Ref.~\onlinecite{Maeter2012}. The broad lines are visual guides.}
\label{fig:phase_diag_NQR}
\end{figure*}

\subsection{Phase diagram}
\label{str:phase_diag_subsec}

The phase diagram for all three rare earths is rebuilt in a systematic way as a function of \xNQRxFSWT, as shown on Fig.~\ref{fig:phase_diag_NQR}.
$T_c$ data for La-based samples have been scaled so that the maximal transition temperature $T_c^{max}$(La)$=$26~K coincides with $T_c^{max}$(Sm)$=$55~K \cite{Ren2008c}.
For cerium, the scaling of $T_c$ data is done so that there is good overlap with the data for Sm-based samples.
The characterization of our samples is taken almost exclusively from references \cite{Kondrat2009,Hess2009,Klingeler2010,Luetkens2009,Maeter2012,Hammerath2013}, in particular the work of Maeter \etal\ where first estimates of \xNQR\ were used for Ce-based samples.
For samples from the literature, the transition temperatures are taken from the original papers.
For a given technique, the data was harmonized whenever possible by extracting the transition temperature in a systematic way (e.g.: midpoint of the resistivity drop for the superconducting transition).
In any case, the use of different techniques for a given phase transition necessarily yields some additional point dispersion, which is taken into account by the visual guides of Fig.~\ref{fig:phase_diag_NQR}.
The transition lines are well-defined, which shows the adequacy of our NQR approach to the doping.
This is especially true for La-based compounds, with twelve samples from other NQR groups with different samples sources being successfully combined to our eight samples.

The tetragonal-to-orthorhombic structural transition and the transition to long-range-ordered (LRO) magnetism can be described by single $T_S$ and $T_{N,LRO}$ lines for cerium and samarium.
In-between these two transitions, a single $T_{N,SRO}$ line describes the transition to short-range-ordered (SRO) magnetism, which extends slightly into the superconductivity region.
A good scaling of the Ce and Sm $T_c$ data can be achieved, with a scaling ratio $\approx$ 0.73 which is similar to the ratio of the maximal transition temperatures (43$/$55 $\approx$ 0.78) \cite{Maeter2012,Ren2008c}.
While no data is available for highly-doped Sm-based samples, this is evidence that the superconducting regions for Ce and Sm are likely very similar, with maximal $T_c$ being reached only at dopings far above the range where static magnetism and superconductivity meet.
Finally, static magnetism of the rare earths also extends far beyond this range.
Thus, the phase diagrams for these two magnetic rare earths show no sign of differences after rescaling $T_c$.

In the case of lanthanum, several differences occur.
$T_{N,LRO}$ would appear to decrease faster: while few samples show this transition, there are several low-doped superconducting samples which do not show it, therefore constraining it to below \xNQRxFSWT$\approx$0.4.
SRO magnetism is completely absent from superconducting samples, although it cannot be ruled out in magnetic samples as a precursor of the LRO magnetic transition.
Finally, the superconductivity region starts at lower doping and grows only until \xNQRxFSWT$\approx$1, after which it decreases with a sign of recovery only at very high doping.
A candidate for the difference in static magnetism compared to the unified behavior of Ce- and Sm-based samples is the lack of magnetism of lanthanum, whose formal 3+ charge implies closed electronic shells.
This is verified experimentally by spin-lattice relaxation rate measurements: the spin fluctuations in La-based samples are much smaller (as well as more doping-dependent) than in samples containing a magnetic rare earth (Ce, Sm, Nd), and are thus ascribed to iron plane magnetism for the former and to $4f$ rare earth magnetism for the latter \cite{Lang2010,Yamashita2010a,Rybicki2012,Ahilan2008,Prando2010}. 
Static magnetism in the iron planes may thus be promoted at higher doping by magnetic rare earths (to be discussed in Sec.~\ref{str:magnetism}), and the concomitant delaying of superconductivity is a strong indication that static magnetism is detrimental to superconductivity.
Regarding superconductivity at high fluorine doping in La-based samples, while it was already known that $T_c$ tends to decrease \cite{Hess2009}, our phase diagram shows to our knowledge the first indication of eventual $T_c$ recovery. 
This is compatible with observations on hydrogen-doped samples \cite{Iimura2012}: a second superconductivity dome occurs for La at high doping, whereas for Ce and Sm there is a single dome extending to more than 40\% doping.
This difference is then argued to be related to the doping-dependence of the iron $3d$ orbital degeneracy.

Note that the ``ribbons'' of Fig.~\ref{fig:phase_diag_NQR} should be seen as an interval of confidence for each transition line.
Especially, the T$\rightarrow$0 superposition of the magnetic and superconducting phase transitions about \xNQRxFSWT$\approx$0.4 (La) and $\approx$0.5 (Ce, Sm) should not be understood as a suggestion of a quantum critical point, whose relevance is anyhow unclear in an inhomogeneous context.
There may be a crossing of the two transitions lines, which must however be very limited in doping range considering that no sample is known to feature both long-range magnetic order and superconductivity.
This will be discussed in terms of ground state competition in Sec.~\ref{str:interplay}.

\begin{figure*}[ht]
\includegraphics[width=130mm]{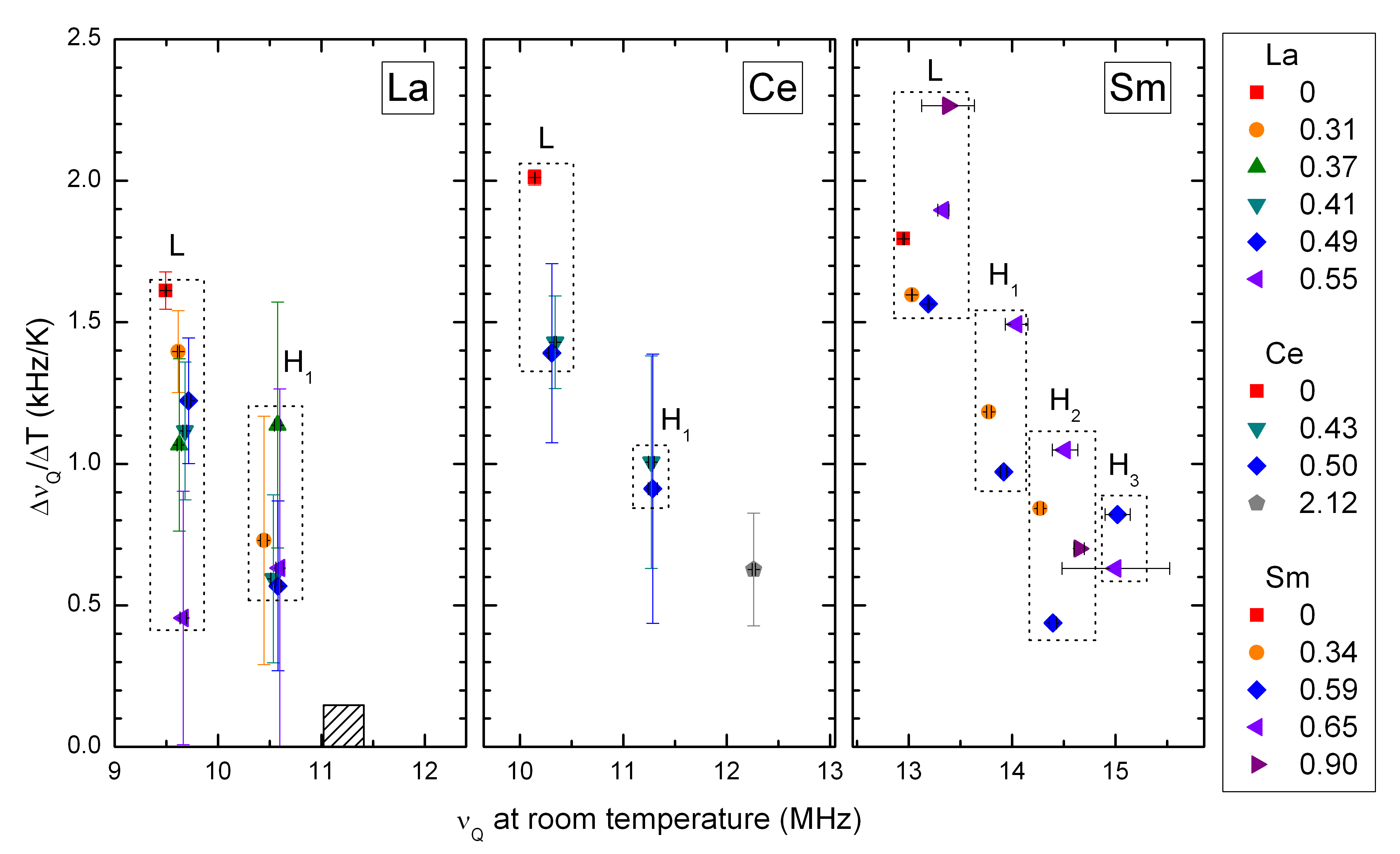}
\caption{(Color online) Temperature sensitivity of the quadrupolar frequency, as a function of the peak position at room temperature. $\Delta \nu_Q / \Delta T$ is calculated from measurements at room temperature and at $T$$=$150--160~K.
The legend indicates \xNQRxFSWT.
For \xNQR$<$\xFSWT, each dotted box groups the points belonging to the same frequency branch (L, H$_1$, H$_2$, H$_3$) as defined on Fig.~\ref{fig:nuQvsXnqr}.
For La-based samples, the hatched rectangle about $\nu_Q$$\approx$11.2~MHz indicates the frequency range over which $\Delta \nu_Q / \Delta T$ goes from weak (\xNQRxFSWT$=$1.19) to zero or negative (\xNQRxFSWT$=$1.90).
Error bars for the Sm-based samples are not shown, as they exceed the actual point dispersion for each frequency branch, suggesting over-estimation due to more complex spectra.}
\label{fig:slopeVSfreq}
\end{figure*}

Considering the electronic separation into regions which seem related either to the magnetically-ordered, undoped compound or to the superconducting-only \xNQRxFSWT$=$1 compound (see Fig.~\ref{fig:nuQvsXnqr}), it is a distinct possibility that these regions are respectively more prone to develop static magnetism or superconductivity. Taking into account our observations about the phase diagram and the fact that most samples exhibit only one of the two ground states, we examine this issue in the next two sections.

\section{Orthorhombicity and static magnetism}
\label{str:magnetism}

\subsection{Spatial origin}

As shown on Figs.~\ref{fig:spectra} and \ref{fig:nuQvsXnqr}, the spectra in the paramagnetic state display a sizeable temperature dependence for most of the samples, with an increase of the resonance frequency when increasing the temperature.
The effect is largest for the undoped samples, and appears reduced or even canceled for single-peaked spectra at high doping, such as for the \xNQRxFSWT$=$1.90 La-based sample.
When the spectrum features two peaks, the temperature dependence of the high-frequency peak seems to be lower than that of the low-frequency peak, as can be seen for instance for the \xNQRxFSWT$=$0.65 Sm-based sample.
An increase of $\nu_Q$ with increasing temperature is also observed in other iron pnictides \cite{Kitagawa2008,Baek2011a}, and goes counter to the expected effect of lattice expansion and lattice vibrations \cite{Nishiyama1976}.
This suggests that other changes take place with temperature, for instance intracell atomic displacements, which are beyond the scope of the present study and are left as an open question.

In the following, we focus on the $T$$\approx$160--300~K range, i.e., a temperature range above all low-temperature transitions. 
Using the data plotted in Fig.~\ref{fig:nuQvsXnqr}, the ratio \DnuDT\ is extracted for all spectral peaks of all samples over this temperature range.
It is then plotted on Fig.~\ref{fig:slopeVSfreq} versus the resonance frequency of each corresponding peak at room temperature, i.e., as far away as possible from low-temperature physics.
Such an abscissa allows to reflect the frequency branch structure of Fig.~\ref{fig:nuQvsXnqr} (see dotted boxes on Fig.~\ref{fig:slopeVSfreq}).
While the uncertainty on \DnuDT\ for a given peak (a single point) is large, for a given rare earth the peaks belonging to the same branch feature a dispersion in \DnuDT\ which is small enough to differentiate the various frequency branches, especially for Ce- and Sm-based samples.
The tendency is for \DnuDT\ to decrease when going from the L frequency branch to the H$_1$ branch (and then to the H$_2$/H$_3$ branches for Sm-based samples), whereas for a given branch large differences in the doping and low-temperature properties of the samples appear to be of lesser importance.
This shows that the temperature dependence of the electronic charge environment at each arsenic site is a local electronic property, rather than an emergent property of the whole sample at a given overall doping.
Note how this is different to previously-obtained $1/T_1$ spin-lattice relaxation rate measurements on La-based samples \cite{Lang2010}, which showed a large doping dependence for both L- and H-branch sites.
We ascribe this difference to $\nu_Q$ being sensitive to the charge degree of freedom (a defined configuration of $3d$ orbital occupancies) and to $1/T_1$ probing the spin degree of freedom (doping-dependent magnetic fluctuations). 

\begin{figure}[pbth]
\includegraphics[width=85mm]{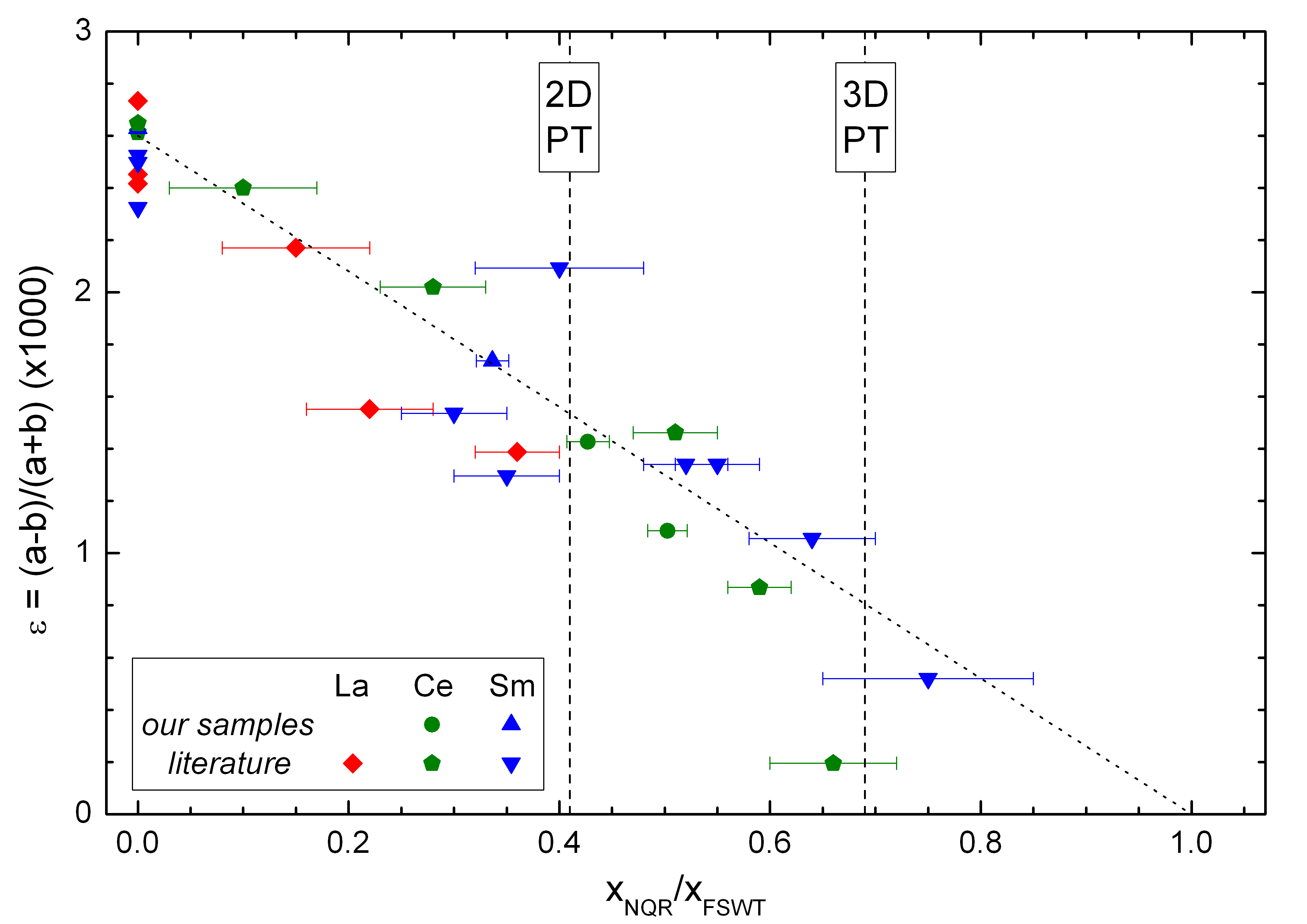}
\caption{(Color online) Order parameter of the structural phase transition $\epsilon = (a-b)/(a+b)$  ($a$, $b$: in-plane lattice parameters) as a function of doping. The structural data for all samples is taken from Refs.~\onlinecite{Maeter2012,Huang2008,Qureshi2010,Zhao2008,Margadonna2009a,Luo2009b,Martinelli2011}. The dotted line is a visual guide. Vertical dashed lines indicate the percolation thresholds for 2D and 3D nearest-neighbor square-lattice site percolation.}
\label{fig:epsilonVSxNQR}
\end{figure}

The fact that the local electronic properties of the undoped, magnetically-ordered compound are well-retained across the L frequency branch suggests that the low-frequency regions may play the key role in establishing orthorhombicity and static magnetism, whereas the high-frequency regions may play no role at all.
To test this hypothesis, the order parameter $\epsilon = (a-b)/(a+b)$ ($a$, $b$: in-plane lattice parameters) of the structural transition is plotted as a function of \xNQRxFSWT\ in Fig.~\ref{fig:epsilonVSxNQR}.
In addition to data from Ref.~\onlinecite{Maeter2012} which corresponds to our samples, for which \xNQR\ is known, we include several other studies from the literature \cite{Huang2008,Qureshi2010,Zhao2008,Margadonna2009a,Luo2009b,Martinelli2011}.
Since the latter do not include NQR data, we use the reported phase transition temperatures to extract \xNQR\ from our reconstructed phase diagram, as shown in Appendix~\ref{str:doping_from_phasediag}.
The agreement between all studies in Fig.~\ref{fig:epsilonVSxNQR} is good, which validates our determination of the doping.
The decrease of $\epsilon$ with \xNQR\ is compatible with a linear decrease until $\epsilon$$=$0 for \xNQR$=$\xFSWT, i.e., when the low-frequency spectral weight vanishes.
This is clear evidence of the low-doping-like regions fully accounting for the orthorhombicity of the material, and thus for the closely-associated static magnetism.

\subsection{Doping-dependence}

From Fig.~\ref{fig:epsilonVSxNQR} and Fig.~\ref{fig:phase_diag_PT}, where all the samples considered in this article are present, it can be seen that \xNQRxFSWT$=$0.40(4) (Oka-0.03, see Tab.~\ref{tab:NQR}) is the highest doping for which a structural or magnetic transition is reported in a La-based sample, and \xNQRxFSWT$=$0.37(4) (one of our samples) the lowest one for which no such transition is observed.
This defines a narrow threshold \xNQRxFSWT$\approx$0.39(3) beyond which only superconductivity occurs.
For Ce samples this threshold is about 0.66(6) (Zhao-0.1, for which vanishing orthorhombicity is reported).
For Sm samples it is at 0.67(2), i.e., between 0.75(10) (Martinelli-0.2) and 0.65(4) (one of our samples).
Such values agree well with the known percolation thresholds for the square-lattice site problem in two and three dimensions, which given as dilution thresholds (i.e., the fraction of sites to remove) are respectively close to 0.41 and 0.69 \cite{Frisch1961,Sykes1964,Newman2000}.
This strongly suggests that the disappearance of orthorhombicity and static magnetism on doping is ultimately limited by the dilution of low-doping-like iron atoms, without influence of the superconductivity.
Note that the interpretation of the above values could be complicated by the fact that each As atom probes a square of four Fe atoms, including a minority of As atoms at the boundaries between \LDL\ and \HDL\ regions a few nanometers in size (see Sec.~\ref{str:SC}).
However, the linear variation of \xNQRxFSWT\ from 0 to 1 upon doping indicates that it is indeed equal to the relative weight of ``\xNQR$=$\xFSWT''-like iron atoms.

The change from two-dimensionality (La) to three-dimensionality (Ce, Sm) indicates an enhanced coupling between the iron planes when the rare earth is Ce or Sm.
Structural effects can be ruled out since there are only marginal differences, especially between La- and Ce-based compounds \cite{Nitsche2010}.
A more likely explanation is the magnetism of the rare earth (Ce, Sm) itself, whose $4f$ magnetic moment should increase interplane coupling compared to the nonmagnetic rare earth (La).
Indeed, magnetic rare earths are known to couple to the $3d$ electrons \cite{Jeglic2009,Maeter2009,Alfonsov2010,Prando2010,Stockert2012}, and are sufficiently coupled to each other to develop an ordered magnetic state at a few kelvins.
Note also that they should not significantly alter the in-plane couplings, as they lie directly above/below iron atoms.
Such a situation is compatible with the parent compound having similar $T_N$ for all rare earths.
Indeed, for the small ratio $r = J_c / J_{ab}$ of the interplane to the intraplane coupling in 1111 compounds \cite{Fang2008a,Ramazanoglu2013}, $T_N$ is predicted to vary little with $r$.
As an example, assuming localized spins $S$$=$1/2, a tripling of $r$ from 0.001 to 0.003 yields only an increase of $T_N$ by 13\% \cite{Yasuda2005}.
In this picture of increased interplane coupling, it is interesting to note that the short-range-ordered (SRO) magnetism transition line of Ce and Sm samples shows a change in curvature between the dopings 0.34(2) and 0.50(2), i.e., about the 2D percolation threshold (see Fig.~\ref{fig:phase_diag_NQR}).
On crossing the threshold, the in-plane density of low-doping-like regions becomes low enough that in-plane couplings between these regions is weakened, effectively increasing the dimensionality. 

In light of the structure, it is expected that in-plane magnetic couplings occur between nearest-neighbor (NN) and next-nearest-neighbor (NNN) iron sites, i.e., that the $J_1$--$J_2$ model applies.
Square lattice NN+NNN percolation has been argued to occur for La/Sm/Pr-1111 with ruthenium substitutions at iron sites \cite{Bonfa2011,Sanna2011,Yiu2014}.
There, spin dilution would cause static magnetism to vanish for $x_{Ru}$$\approx$0.6, close to the theoretical value 0.593 \cite{Malarz2005}.
However, there is significant evidence of a nonrandom distribution of Ru and microscopic phase separation in Ru-poor and Ru-rich regions \cite{Sanna2011,Iadecola2012,Joseph2013,Martinelli2013,Simonelli2014,Martinelli2014}, suggesting that the global Ru content is not the relevant parameter to describe percolation.
In Appendix~\ref{str:phase_sep_Ru}, we propose a simple model to reinterpret the NQR data of Sanna \etal\ \cite{Sanna2011} by assuming a phase separation in Fe-rich and Ru-rich regions.
For the $x_{Ru}^{nominal}$$=$56\% composition, close to which static magnetism and superconductivity vanish, we find a volume fraction $\approx$48\% of Ru-rich regions with 80\% local Ru concentration, which is likely too high to allow for any static magnetism of the iron atoms embedded in the nonmagnetic Ru matrix \cite{Tropeano2010}.
Static iron magnetism should instead originate from the $\approx$52\% volume fraction of Fe-rich regions, which harbor low-doping-like (\LDL) areas as defined in the present article.
Due to the phase separation, the Ru-rich regions should be inefficient at preventing percolation of these \LDL\ areas.
It is thus striking that the fraction of \LDL\ areas in the Fe-rich regions is equal to 0.274(66), i.e., consistent with 3D NN percolation (threshold equal to 0.31) rather than with 2D NN+NNN percolation (0.407).
Although further study of the role of Ru substitutions is needed, this suggests agreement with the behavior observed by us in F-doped samples with a magnetic rare earth.
Note that comparison with LaFeRuAsO using the NQR data of Ref.~\onlinecite{Mazzani2014} could not be performed, since the very broad magnetic transitions for La-based samples \cite{Bonfa2011,Sanna2011,Yiu2014} imply a more complicated distribution of ruthenium in the iron planes, for instance a variability of the ruthenium content in the iron-rich regions of each sample.
One possibility for the observation of NN percolation is that the elementary regions which percolate to yield static magnetism are larger than one lattice cell and interact through some effective NN interaction, whether local in character or involving the conduction electrons.
Besides, even in the site-to-site $J_1$--$J_2$ picture, complications are expected, such as the NNN hoppings eventually dominating the NN hoppings at high doping \cite{Suzuki2014}.

\subsection{Mechanism}

An immediate consequence of the observed percolation behavior is that the loss of static magnetism when doping is not related to a degradation of the nesting properties of the Fermi surface.
An even more fundamental question about iron pnictides is whether the tightly-related structural and magnetic transitions are driven by orbital or spin fluctuations \cite{Fernandes2014}.
In the presence of a magnetic rare earth, not only is 3D magnetism obtained on doping beyond the 2D percolation threshold, but orthorhombicity also occurs until the 3D percolation threshold (seemingly with an increased dimensionality \cite{Maeter2012}).
Considering the crystal structure where $R$-(O,F)-$R$ ($R$: rare earth) layers separate the FeAs layers, it is hard to explain how the unpaired $4f$ electrons of the rare earth would promote orbital correlations, rather than spin correlations.
This suggests that the electronic anisotropy of the 1111 system at low doping is primarily driven by spin fluctuations, eventually yielding the enhanced, slow spin-nematic fluctuations seen below the structural transition \cite{Fu2012,Zhang2015}, or at least that there is a decisive feedback of spin fluctuations on orbital fluctuations.

\section{Superconductivity}
\label{str:SC}

The high-frequency, high-doping-like regions are natural candidates to harbor superconductivity.
However, the superconducting volume fraction of 1111 materials appears to be large even at low doping \cite{Kondrat2009}.
Previous spin-lattice relaxation measurements also showed that the low- and high-frequency regions show a similar decrease of spin excitations below $T_c$ \cite{Lang2010,Oka2011}.
These observations could suggest that the low-frequency, low-doping-like regions support superconductivity when static magnetism is repressed, which is a crucial point with respect to ground state competition and coexistence.

\begin{figure}[pbth]
\includegraphics[width=85mm]{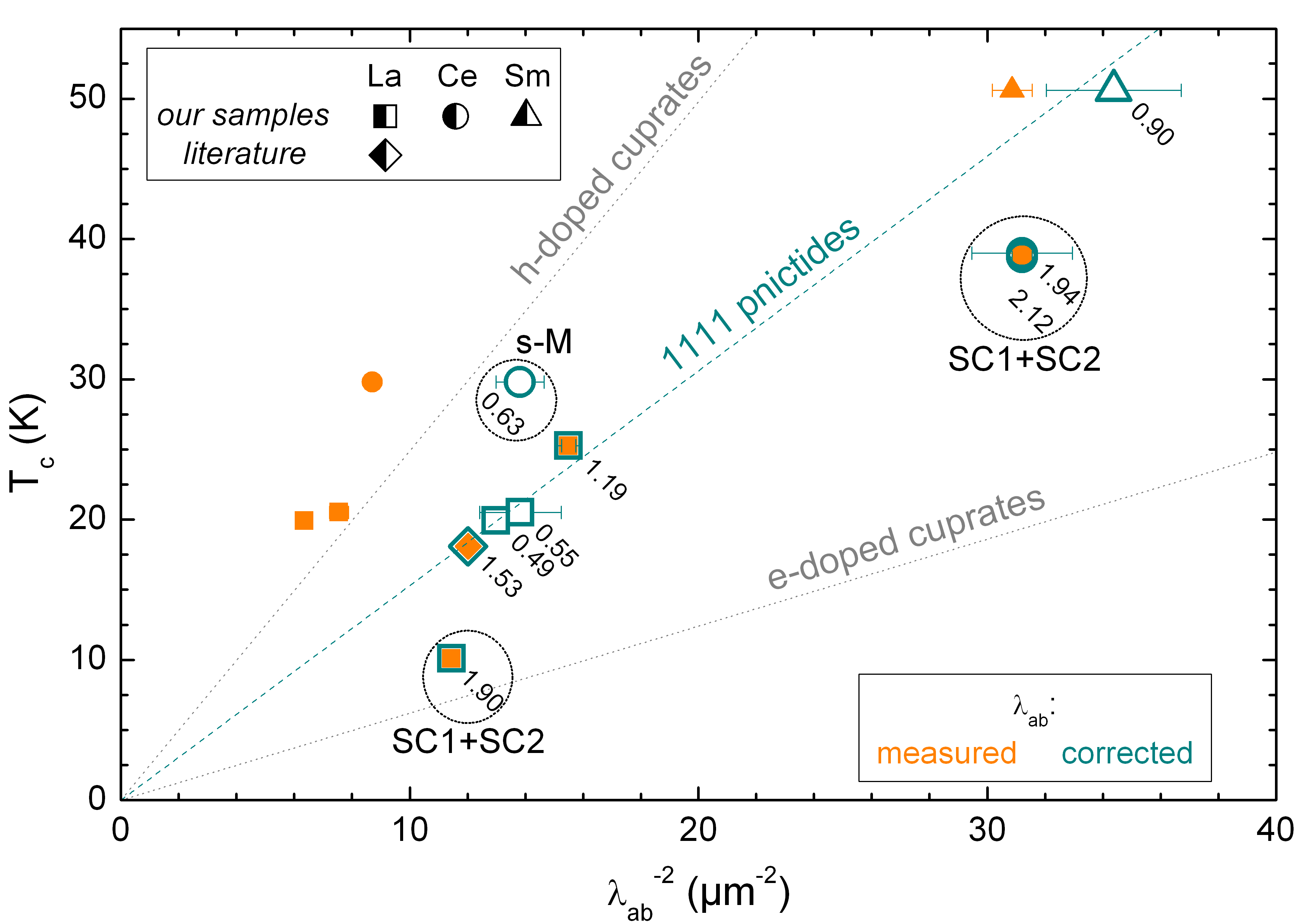}
\caption{(Color online) Uemura plot showing the critical temperature versus $\lambda_{ab}^{-2}$. All measured data is taken from Refs.~\onlinecite{Luetkens2008,Luetkens2009,Maeter2012} (including one sample whose \xNQRxFSWT\ is determined from its $T_c$ value, see Appendix~\ref{str:doping_from_phasediag}). \xNQRxFSWT\ is shown as a label. The uncertainty on $\lambda_{ab}^{-2}$ is given for all samples where uncertainties on $\lambda_{ab}$ and \xNQR\ are both available. The dashed line is a visual guide. Dotted lines indicate the behavior observed for hole-doped and electron-doped cuprates below or at optimal doping \cite{Uemura1991,Shengelaya2005}. ``s-M'' refers to the presence of strong static magnetism, and ``SC1+SC2'' to the crossover between two superconducting states.}
\label{fig:Uemura}
\end{figure}

In La-1111, there is a significant decrease of the superfluid density $\rho_s$ below optimal doping \cite{Luetkens2009}.
In the case where both ground states would coexist, competition for carriers would be expected.
Such is the case in the 122 compound Ba(Fe$_{1-x}$Co$_x$)$_2$As$_2$, where a large enhancement of the penetration depth $\lambda_{ab}$ is observed when the doping is low enough for coexistence \cite{Gordon2010a}.
$\lambda_{ab}^2$ is inversely proportional to $\rho_s$, whose decrease is argued to be due to the SDW partially gapping the Fermi surface, i.e., removing carriers.
Even though no static magnetism is present in superconducting La-1111, it cannot be ruled out that the remaining slow spin fluctuations seen by NMR below optimal doping \cite{Hammerath2013} are related to a partial gapping of the Fermi surface over some volume fraction of the samples.
However, the observed decrease of $\rho_s$ is much faster than that of $T_c$, yielding a Uemura plot (closed orange symbols on Fig.~\ref{fig:Uemura}) where the pnictides show no well-defined slope as opposed to underdoped cuprates \cite{Uemura1991}.
Taken at face value, this would suggest that superconductivity presents large variations of the coupling strength versus doping.
Writing $\rho_s$$=$$n_s/m^*$ with $n_s$ the superfluid carrier density and $m^*$ the effective mass of the Cooper pairs, and considering the complex multiband structure, an explanation would be that changes in $m^*$ occur with doping \cite{Hammerath2014}.

Here, we argue that the measured superfluid density (usually using $\mu$SR) is reduced for \xNQRxFSWT$<$1 due to the proximity effect.
Indeed, if only the \HDL\ regions are intrinsically superconducting, then their nanoscale coexistence with the \LDL\ regions implies that the proximity effect may play over a large volume fraction, up to the entirety of the \LDL\ regions.
Assuming the extremal case, the measured superfluid density then corresponds to a 100\% SC volume fraction originating only from the \HDL\ regions, i.e.:
\begin{equation}
\rho_s^{m} = \rho_s^{i} \frac{\xNQR}{\xFSWT}   \mbox{,}
\label{eq:superfluid}
\end{equation}
where $\rho_s^{m/i}$ are the measured and intrinsic superfluid densities.
Applying this correction to our samples for which penetration depth data is available \cite{Luetkens2008,Luetkens2009,Maeter2012}, this yields a well-defined $T_c$$=$$f(n_s/m^*)$ slope, in-between that of hole- and electron-doped cuprates, for the La samples with 0.49$\le$\xNQRxFSWT$\le$1.53 and for the Sm sample with \xNQRxFSWT$=$0.90 (open turquoise symbols and visual guide on Fig.~\ref{fig:Uemura}).
This single slope indicates a common, well-defined superconducting state behavior, suggesting the validity of this scenario where only the \HDL\ regions intrinsically support superconductivity.
Even in the absence of static magnetism, the \LDL\ regions do not appear to contribute to the condensate.
This suggests that in the undoped limit there is no underlying superconducting ground state which static magnetism would be suppressing.
The implications for the mechanism responsible for Cooper pairing are also strong, since the linearity of $T_c$ on $\rho_s$ would seem to rule out theoretical approaches in the weak-coupling limit \cite{Uemura1988}.

Note that, for dopings where the material is electronically inhomogeneous (\xNQRxFSWT$<$1), it cannot be said that the observed behavior is that of a material that is simply less doped than \xFSWT.
The observed Uemura relation then rather reflects the properties of \xFSWT-like superconductivity which is weakened due to the interspersed low-doping-like and high-doping-like regions.
This seems however to be equivalent to a doping variation, since the electronically-homogeneous, roughly optimally-doped \xNQRxFSWT$=$1.19 La-based sample obeys the same behavior.
For La samples with even higher \xNQRxFSWT, whose $T_c$ would suggest they are overdoped, the Uemura relation appears to be maintained through the concomitant decrease of $T_c$ and $\rho_s$ up to at least \xNQRxFSWT$=$1.53, and breaks for \xNQRxFSWT$=$1.90 which shows a reduced $T_c$ and a still moderate superfluid density.
This is unlike the behavior observed in cuprates, where overdoping tends to result in a saturation then a decrease of $T_c$ as $\rho_s$ keeps increasing \cite{Uemura1991}.
Here, the \xNQRxFSWT$=$1.53 La-based sample is behaving like an underdoped sample, suggesting that the observed $T_c$ dome is misleading.
Note that this is in agreement with the study of hydrogen-doped 1111 pnictides by Iimura \etal\ \cite{Iimura2012}, who report that the decrease of $T_c$ beyond optimal doping may be due to a degradation of the nesting of the Fermi surface and of the associated spin fluctuations.
Concerning the \xNQRxFSWT$=$1.90 La-based sample, it is interesting to note that the breakdown of the Uemura relation occurs in the region of the phase diagram where $T_c$ is depressed (see Fig.~\ref{fig:phase_diag_NQR}), which corresponds to a crossover between two different superconducting states (SC1 and SC2) \cite{Iimura2012}.
Assuming that SC1 and SC2 may be competing but do not support each other, we define in the following $T_c^{(1),(2)}$ as their respective critical temperatures at a given doping, and $\rho_s^{(1),(2)}$ as their respective contributions to the total superfluid density $\rho_s$.
For \xNQRxFSWT$=$1.90, Fig.~\ref{fig:phase_diag_NQR} indicates $T_c^{(1)}$$>$$T_c^{(2)}$.
While the available data is insufficient to conclude, a possible scenario is then that the Uemura relation is still obeyed for SC1 with $T_c$$=$$T_c^{(1)}$, and with $\rho_s^{(2)}$ accounting for the apparent breakdown.
For highly-doped Ce-based samples, the breakdown of the Uemura relation could have a similar explanation.
Finally, the insufficient correction of the superfluid density observed for the \xNQRxFSWT$=$0.63 Ce-based sample will be linked to ground state competition in Sec.~\ref{str:interplay}.

Our result on the presence of a full-volume proximity effect allows to put an upper boundary $\xi_N$ on the typical distance $d_{H-H}$ separating \HDL\ regions, with $\xi_N$ the coherence length in the normal material, i.e., the characteristic length over which the pair amplitude decays.
For similarly poor conductors such as cuprates, $\xi_N$ can be expected to be on the order of a few nanometers.
Since the \LDL\ regions are unable to support superconductivity by themselves, it is possible to rule out an enhancement of $\xi_N$ at low temperature such as that observed in Josephson junctions where two superconducting, high-$T_c$ cuprate electrodes are separated by an underdoped, lower-$T_c$ cuprate barrier in the normal state \cite{Bozovic2004,Kirzhner2014}.
The small value of $d_{H-H}$, together with the conducting electronic background, favors the strong coupling of the intrinsically-superconducting regions \cite{Tanabe2011}.
This may explain why, in the absence of static magnetism, a sample with \xNQRxFSWT\ as low as 0.49 (La-based sample) still achieves a $T_c$ as high as $\approx$75\% of the optimal value.
Such a strong coupling was also proposed to explain the resilience of superconductivity in Ru-doped Ba-122 \cite{Laplace2012}.
Finally, the high-doping-like regions should be at least of size $\xi_S$, the superconducting coherence length.
This translates to about 20--40~\AA\ in the $ab$ plane \cite{Sefat2008,Pallecchi2009}.
Therefore, a patchwork of regions a few nanometers across is the most simple arrangement compatible with the observed superconducting behavior.
A similar conclusion was reached by Sanna \etal\ on the basis of the magnetic behavior \cite{Sanna2009,Sanna2010}.

\section{Interplay of the ground states}
\label{str:interplay}

In light of the detrimental effect of iron static magnetism on superconductivity (see Sec.~\ref{str:phase_diag_subsec}), ground state coexistence appears as a special case where static magnetism from the \LDL\ regions may restrict the spatial extension of superconductivity originating from the \HDL\ regions.
Indeed, such a behavior seems present in the \xNQRxFSWT$=$0.63 Ce-based sample.
As seen in Fig.~\ref{fig:Uemura}, correcting the superfluid density of this sample according to Eq.~\ref{eq:superfluid} is not enough to bring it in agreement with the observed Uemura relation.
This suggests a loss of superfluid density, on contrary to the samples without ground-state coexistence.
According to $\mu$SR the magnetic volume fraction reaches 100\%, with a freezing of the iron moments over an extended temperature range ($\approx$5--26~K) \cite{Maeter2012}.
Such a distribution of $T_N$ is a common feature \cite{Sanna2009,Sanna2010,Shiroka2011,Lamura2015}.
The 100\% volume fraction does not mean that the whole sample is intrinsically magnetic, which would be incompatible with our results, but that the field from the magnetically-ordered \LDL\ regions is also felt in the \HDL\ regions due to the nanoscale separation. 
At a given $T_N$, the corresponding magnetic regions and their nanometer-scale environment will show up as frozen volume in the $\mu$SR measurement.
In isovalently-doped Ba(Fe$_{1-x}$Ru$_x$)$_2$As$_2$, the distribution of $T_N$ was ascribed to a spatial variation of the ordered iron moment, with values smaller than about 0.3~$\mu_B$ allowing for local coexistence of superconductivity and static magnetism \cite{Laplace2012}.
In F-doped 1111 pnictides, the variation of the strength of static magnetism could be due to finite-size effects and to the disorder inherent to a transition with percolative character (see Sec.~\ref{str:magnetism}).
The \LDL\ regions with the strongest magnetism (highest $T_N$) will locally suppress superconductivity by proximity, and possibly also the intrinsic superconductivity in the nearest \HDL\ regions, whereas the \LDL\ regions with weak or zero static magnetism will tolerate it.
In such a picture some superfluid density is irremediably lost in or close to the more magnetic regions, in agreement with the observed breakdown of the Uemura relation.
This would also explain the report of a reduced superconducting volume fraction in the presence of ground state coexistence \cite{Sanna2009}.

\begin{figure*}[htbp]
\includegraphics[width=175mm]{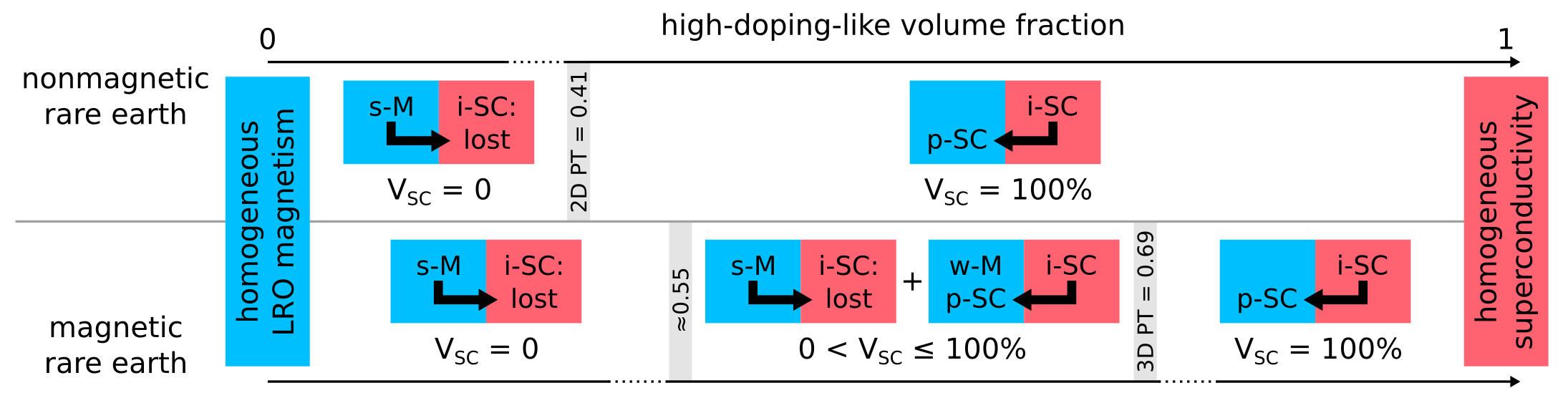}
\caption{(Color online) Summary of the interplay of ground states in the doping range where \LDL\ (blue) and \HDL\ (red) regions coexist at the nanoscale. The \LDL\ regions can feature strong (``s-M'') or weak (``w-M'') static magnetism, as well as superconductivity through proximity (``p-SC''). The \HDL\ regions feature intrinsic superconductivity (``i-SC''), which can be suppressed (``i-SC: lost''). \VSC\ is the superconducting volume fraction, while ``LRO'' and ``PT'' stand for ``long-range ordered'' and ``percolation threshold''. For the magnetic rare earth case, the intermediate doping region (0$<$\VSC$\le$100\%) is simplified by not making a distinction between strong static magnetism preventing superconductivity by proximity and destroying nearby intrinsic superconductivity, and by not detailing the behavior close to the 3D percolation threshold, where all of the static magnetism is expected to be weak enough for \VSC$\approx$100\%.}
\label{fig:GSinterplay}
\end{figure*}

All our results which pertain to the existence and the interplay of static magnetism and superconductivity are summarized in Fig.~\ref{fig:GSinterplay}.
In the parent compound, the paramagnetic state is electronically homogeneous, yielding the well-known long-range-ordered magnetic phase at low temperature.
On increasing the doping nonmagnetic \HDL\ regions start to appear, while $T_N$ decreases. 
The static magnetism from the \LDL\ regions is nonetheless still strong enough to suppress any superconductivity that may have emerged from the \HDL\ regions, thus a zero superconducting volume fraction (\VSC).
For the nonmagnetic rare earth (La), orthorhombicity and static magnetism cannot occur beyond the 2D nearest-neighbor percolation threshold.
The development of superconductivity is then no longer hampered anywhere in the material, and \VSC\ jumps to 100\% due to superconductivity by proximity in the \LDL\ regions.
A homogeneous, purely intrinsic superconducting state is eventually reached when the \LDL\ regions have disappeared.
For the magnetic rare earths (Ce, Sm) static magnetism survives to higher doping, allowing for a distribution of $T_N$ that crosses progressively into a low-temperature range: the \LDL\ regions with weaker magnetism are no longer able to suppress superconductivity originating from the nearby \HDL\ regions, and \VSC\ starts to grow.
This occurs at \xNQRxFSWT$\approx$0.55, i.e., in-between 0.50 (purely magnetic Ce sample) and 0.59 (magnetic and superconducting Sm sample), with the approximation that the boundary is the same for Ce- and Sm-based compounds.
On approaching the 3D nearest-neighbor percolation threshold (\xNQRxFSWT$\approx$0.69), $T_N$ should be systematically low before actually vanishing.
Intrinsic superconductivity in the \HDL\ regions is then never suppressed, nor is superconductivity by proximity in the \LDL\ regions.
Again, a homogeneous superconducting state is reached for \xNQR$=$\xFSWT.
While this scenario accounts well for the qualitative difference between the phase diagrams of nonmagnetic and magnetic rare earths \cite{Luetkens2009,Drew2009,Sanna2009,Sanna2010}, note that a small coexistence region cannot be ruled out for La-based samples, as suggested by measurements on hydrogen-doped samples \cite{Lamura2014}.
Any such coexistence region will be limited to the very narrow doping range (\xNQRxFSWT$=$0.39$\pm$0.03, see Sec.~\ref{str:magnetism}) over which static magnetism could be weakened enough to allow for the presence of superconductivity.
It should also be noted that for Ce- and Sm-based samples at intermediate doping, the detailed evolution of \VSC\ should depend not only on the distribution in the strength of the static magnetism but also on the microscopic arrangement of \LDL\ and \HDL\ regions, which will affect how much superconductivity is suppressed.

\section{Conclusion}

The phase diagram of 1111 $R$FeAsO$_{1-x}$F$_x$ ($R$$=$La, Ce, Sm) iron pnictides has been constructed as a function of the NQR spectral properties in the paramagnetic state, i.e., the properties of two types of local charge distributions associated to low-doping-like (LD-like) and high-doping-like (HD-like) regions.
The combination of a local probe technique and relative intensity measurements results in high accuracy and consistency across all measured samples.
This allows to show without any ambiguity that magnetic rare earths (Ce, Sm) promote iron static magnetism to higher doping, with a detrimental effect on superconductivity.
Besides, our approach allows for successful inclusion of NQR data from other groups with different sample sources, yielding to our knowledge the most extensive phase diagram for fluorine-doped La-based compounds, including the upturn of the superconducting transition temperature at high doping.

Using the NQR-defined doping, as directly measured for our samples or as derived from our phase diagram for studies using other techniques, we have investigated the spatial origin of static magnetism and superconductivity.
It is found that the \LDL\ regions are closely associated to the development of orthorhombicity and static magnetism, with the upper doping limit set by dilution effects: 2D (respectively 3D) nearest-neighbor square lattice site percolation is at play when the rare earth is nonmagnetic (respectively magnetic).
The \LDL\ regions are not intrinsically supportive of superconductivity but can harbor superconductivity by proximity, originating from the nearby \HDL\ regions, whenever static magnetism is weak enough.
In the end, the interplay of ground states in 1111 pnictides appears to be well described as a spatial competition between nanoscopic regions, which compete to establish the ground state through suppression of superconductivity by static magnetism, and extension of superconductivity by proximity effect.

Our conclusions are compatible with various observations taken from the literature, whether associated to static magnetism, such as a pseudogap-like feature that survives up to $T$$\approx$140~K in superconducting samples and the Nernst signature of a possible SDW precursor in an underdoped superconducting sample without static magnetism \cite{Gonnelli2009a,Kondrat2010}, or associated to superconductivity, such as the presence of additional pinning due to nanoscale regions with suppressed superconductivity and the possibly-related enhancement of the upper critical field $H_{c2}$ and $dH_{c2}/dT(T_c)$ at low doping \cite{Panarina2010,Kohama2009}.
The well-defined Uemura relation is also in agreement with work supporting the strong coupling nature of superconductivity in iron pnictides \cite{Lee2009}.
Besides, our scenario eliminates the controversy surrounding the ``first-order-like'' transition in the $R$=La phase diagram from magnetically-ordered samples to superconducting samples. Rather than a smooth evolution of the ground state throughout the whole sample volume, the conditions for magnetic order or superconductivity are already met on a local scale, with the switching of the ground state being dependent on percolation.

Through a better understanding of ground state competition and of the nature of superconductivity, our results should help bring insight on the origin of unconventional superconductivity in iron pnictides.
Further work should address the origin and the details of the nanoscopic arrangement of magnetism-prone and superconductivity-prone regions, for which some theory leads exist \cite{Luo2011,Misawa2014}, as well as the applicability of our picture of ground state interplay to other iron pnictide families.

\section*{Acknowledgments}

We acknowledge insightful discussions with N. Bergeal, B. B\"{u}chner, S. Caprara, L. de' Medici, A. Descamps-Mandine, B. Fauqu\'{e}, C. Feuillet-Palma, J. E. Hamann-Borrero, J. Jeanneau, A. Trokiner, as well as experimental support from R. M\"uller, K. Leger, C. Malbrich, and M. Deutschmann (synthesis), S. M\"uller-Litvanyi and G. Kreutzer (SEM-WDX), C. G. F. Blum (XRD), O. Vakaliuk (resistivity), S. Ga\ss\ and R. Vogel (NQR).
This work was supported by the Deutsche Forschungsgemeinschaft (DFG) through FOR 538, SPP 1458 (Grants No.~BE1749/13 and No.~GR3330/2), and the Emmy Noether programme (Grant No.~WU595/3-1).

\textsuperscript{*}guillaume.lang@espci.fr

\appendix

\section{Doping determinations using the reconstructed phase diagram}
\label{str:doping_from_phasediag}

\begin{figure*}[htbp]
\includegraphics[width=150mm]{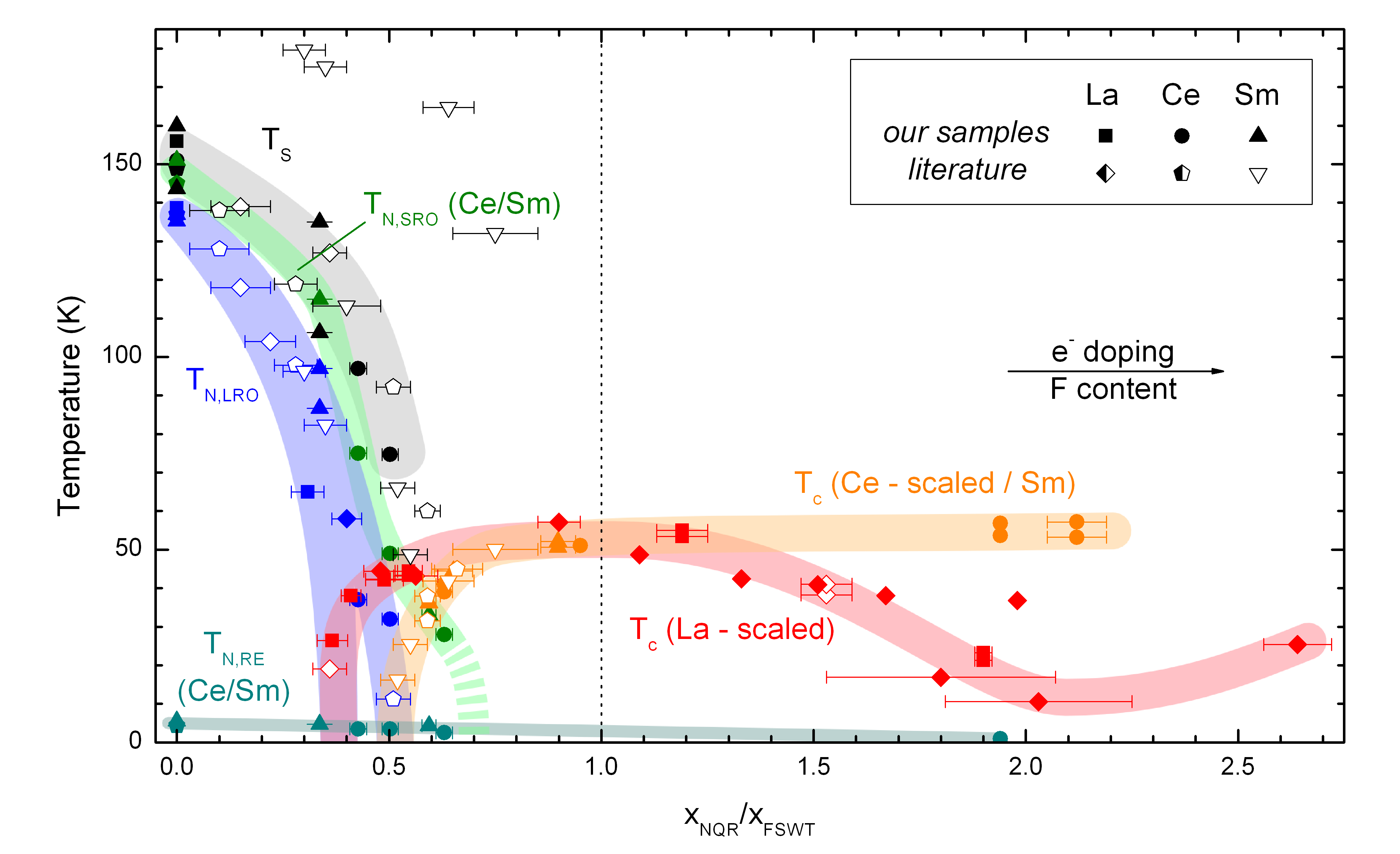}
\caption{(Color online) Phase diagram as used for the determination of \xNQRxFSWT\ for samples from the literature, whose phase transition temperatures are known but for which no NQR data is available (open symbols). These samples are listed in Tab.~\ref{tab:literature_PT}. The closed symbols correspond to the samples used to determine the phase diagram (see also Fig.~\ref{fig:phase_diag_NQR}).}
\label{fig:phase_diag_PT}
\end{figure*}

Using the phase diagram obtained in Sec.~\ref{str:phasediag}, it is possible to extract \xNQRxFSWT\ for samples from the literature whose phase transition temperatures are known, without the need for NQR data.
As shown on Fig.~\ref{fig:phase_diag_PT}, this has been done for a number of samples (listed in Tab.~\ref{tab:literature_PT}) whose properties (orthorhombicity, penetration depth) are of interest for Sec.~\ref{str:magnetism} and Sec.~\ref{str:SC}.
Agreement could be achieved for samples showing more than one transition, with the exception of the samples from Ref.~\onlinecite{Martinelli2011}.
The latter feature structural phase transition temperatures which are unusually high, due to a specific analysis of synchrotron data which is sensitive to incipient order.
For these samples, priority in setting \xNQR\ is then given to the magnetic and superconducting phase transitions.

\begin{table}
\begin{tabular}[c]{cccc}
\hline
\hline
Rare earth	&	Source - first author			& 	$x$	&	\xNQRxFSWT		\\
\hline

La      &       \cite{Luetkens2009} - Luetkens  &       0.125   &       1.53(6)\phantom{1}      \\ 
        &       \cite{Huang2008} - Huang        &       0.03    &       0.15(7)\phantom{1}      \\ 
        &                                                       &       0.05    &       0.36(4)\phantom{1}      \\ 
        &       \cite{Qureshi2010} - Qureshi    &       0.045   &       0.22(6)\phantom{1}      \\ 
 & & & \\ 
Ce      &       \cite{Zhao2008} - Zhao  &       0.02    &       0.10(7)\phantom{1}      \\ 
        &                                                       &       0.04    &       0.28(5)\phantom{1}      \\ 
        &                                                       &       0.06    &       0.51(4)\phantom{1}      \\ 
        &                                                       &       0.08    &       0.59(3)\phantom{1}      \\ 
        &                                                       &       0.1     &       0.66(6)\phantom{1}      \\ 
 & & & \\ 
Sm      &       \cite{Martinelli2011} - Martinelli      &       0.05    &       0.30(5)\phantom{1}      \\ 
        &                                                       &       0.075   &       0.35(5)\phantom{1}      \\ 
        &                                                       &       0.1     &       0.64(6)\phantom{1}      \\ 
        &                                                       &       0.2     &       0.75(10)        \\ 
        &       \cite{Margadonna2009a} - Margadonna     &       0.05    &       0.40(8)\phantom{1}      \\ 
        &                                                       &       0.1     &       0.52(4)\phantom{1}      \\ 
        &                                                       &       0.12    &       0.55(4)\phantom{1}      \\

\hline
\hline
\end{tabular}
\caption{Samples from the literature which are included in Fig.~\ref{fig:epsilonVSxNQR} and Fig.~\ref{fig:Uemura}. $x$ indicates the fluorine doping as given in the source article, which is usually the nominal value. \xNQRxFSWT\ is derived from Fig.~\ref{fig:phase_diag_PT}.}
\label{tab:literature_PT}
\end{table}

\section{Phase separation in Ru-substituted compounds}
\label{str:phase_sep_Ru}

Following the observation of microscopic phase separation in Fe-rich and Ru-rich regions in 1111 samples where iron has been partially substituted with ruthenium \cite{Sanna2011,Iadecola2012,Joseph2013,Martinelli2013,Simonelli2014,Martinelli2014}, we propose a simple model to reinterpret the data of Ref.~\onlinecite{Sanna2011}.
Applying NQR to SmFe$_{1-x}$Ru$_x$AsO$_{0.85}$F$_{0.15}$, Sanna \etal\ obtain spectra featuring low- and high-frequency components like ours but also four more peaks corresponding to As nuclei with 1/2/3/4 Ru nearest-neighbors (NN).
We write the corresponding experimental spectral weights as $w_p^{exp}$ with $p$$=$L/H/1/2/3/4.
For the $x_{Ru}^{nominal}$$=$0.56 compound, in which static magnetism and superconductivity are on the verge of vanishing, NQR yields the weights given in Tab.~\ref{tab:Ru_model} (as well as $w_{H}^{exp}$$=$0).
As then noted there is a tendency towards Ru clustering, with $w_4^{exp}$ being twice larger than expected for a binomial distribution (see $w_p^{bin}$ in Tab.~\ref{tab:Ru_model}).

\begin{table}
\begin{tabular}{cccccc}
\hline
\hline
 	& Experiment	& Binomial	&			& Model			&		\\
\cline{4-6}
 $p$	& $w_p^{exp}$	& $w_p^{bin}$	& $w_p^{i}$		& $w_p^{r}$		& $w_p^{calc}$	\\
 \hline
 L	& 0.142(26)	& 0.037		& 0.274(66)		& 0.001(1)\phantom{0}	& 0.143(26)	\\
 1	& 0.217(26)	& 0.191		& 0.419(10)		& 0.024(13)		& 0.229(20)	\\
 2	& 0.251(26)	& 0.364		& 0.240(47)		& 0.149(45)		& 0.196(24)	\\
 3	& 0.189(26)	& 0.309		& 0.061(25)		& 0.408(22)		& 0.228(26)	\\
 4	& 0.201(26)	& 0.098		& 0.006(4)\phantom{0}	& 0.417(81)		& 0.204(26)	\\
\hline
\hline
\end{tabular}
\caption{Spectral weights for the local environments seen by NQR in SmFe$_{0.44}$Ru$_{0.56}$AsO$_{0.85}$F$_{0.15}$ (see text). The experimental data is taken from Ref.~\onlinecite{Sanna2011}, and the experimental uncertainty is assumed to be the point size of the original figure.}
\label{tab:Ru_model}
\end{table}

Beyond phase separation into Fe-rich and Ru-rich regions, our model is built on these hypotheses: (i) each type of region features binomial statistics corresponding to the local Ru concentration (ii) the phase separation is sufficiently marked that the Fe-rich regions harbor almost all As with zero or one Ru NN, and the Ru-rich regions almost all As with 4 Ru NN.
In the following, all quantities related to the Fe- and Ru-rich regions are respectively indicated by the exponents ``$i$'' and ``$r$''.

We first derive all compositional quantities.
The Ru content in the Fe-rich regions $x_{Ru}^{i}$ is given by:
$$ \frac{(1-x_{Ru}^{i})^4}{4 x_{Ru}^{i} (1-x_{Ru}^{i})^3} \stackrel{\mbox{(i)}}{=} \frac{w_{L}^{i}}{w_1^{i}} \stackrel{\mbox{(ii)}}{\approx} \frac{w_{L}^{exp}}{w_1^{exp}} \mbox{,}$$
where the two equalities derive from hypotheses (i) and (ii).
This yields $x_{Ru}^{i}$$=$0.276(44), thus the $w_{p}^{i}$ weights in Tab.~\ref{tab:Ru_model}.
The volume fraction of Fe-rich regions is given by:
$$ V^{i} \stackrel{\mbox{(ii)}}{\approx} \frac{w_{L}^{exp}}{w_{L}^{i}} = 0.518(69) \mbox{.}$$
The Ru content in the Ru-rich regions $x_{Ru}^{r}$ is given by:
$$ (x_{Ru}^{r})^4 \stackrel{\mbox{(i)}}{=} w_4^{r} \stackrel{\mbox{(ii)}}{\approx} \frac{w_4^{exp}}{( 1 - V^{i})} \mbox{,}$$
yielding $x_{Ru}^{r}$$=$0.804(39), thus the $w_{p}^{r}$ weights in Tab.~\ref{tab:Ru_model}.
The calculated Ru content of the whole sample is given by:
$$ x_{Ru}^{calc} = V^{i} x_{Ru}^{i}  +  (1 - V^{i}) x_{Ru}^{r} = 0.531(22) \mbox{.}$$
The spectral weights for the whole sample are given by:
$$ w_p^{calc} = V^{i} w_p^{i}  +  (1 - V^{i})  w_p^{r}  \mbox{,}$$
with $p$$=$L/1/2/3/4, yielding the $w_{p}^{calc}$ weights in Tab.~\ref{tab:Ru_model}.

We then check the consistency of the calculation.
The consistency of hypothesis (ii) is shown by the following population ratios being small:
$$ \frac{w_{L}^{r} (1-V^{i})}{w_{L}^{i} V^{i}} = 0.005(5) \mbox{,}$$
$$ \frac{w_{1}^{r} (1-V^{i})}{w_{1}^{i} V^{i}} = 0.054(41) \mbox{,}$$
$$ \frac{w_{4}^{i} V^{i}}{w_{4}^{r} (1-V^{i})} = 0.015(11) \mbox{.}$$
This is also reflected in $w_p^{calc}$$\approx$$w_p^{exp}$ for $p$$=$L/1/4, and reasonable agreement is found for $p$$=$2/3.
The small differences should be attributed to the phase boundaries and to minor deviations from randomness inside the Fe-rich and Ru-rich regions.
Finally, the slightly-reduced value of $x_{Ru}^{calc}$ compared to $x_{Ru}^{nominal}$ is in agreement with experimental determinations \cite{McGuire2009,Mazzani2014}.
Therefore, our simple model appears to be in good quantitative agreement with the experiment.
It is likely complementary to that of Ref.~\onlinecite{Martinelli2014}, which cannot easily distinguish the quantitative properties of the iron-rich and ruthenium-rich regions, but provides microscopic arrangements compatible with NQR measurements.

\clearpage

\bibliographystyle{unsrt}

\label{fin}
\end{document}